\documentclass[prl,superscriptaddress,twocolumn,preprintnumbers]{revtex4}
\usepackage{amsfonts}
\usepackage{amsmath}
\usepackage{amssymb}
\usepackage{bm}
\usepackage{dcolumn}
\usepackage{epsfig}
\usepackage{graphicx}
\usepackage{graphics}
\usepackage[latin1]{inputenc}
\usepackage{latexsym}
\usepackage{rotating}
\usepackage{hyperref}

\usepackage{amsfonts}
\usepackage{amsmath}
\usepackage{amssymb}
\usepackage{bm}
\usepackage{dcolumn}
\usepackage{epsfig}
\usepackage{graphicx}
\usepackage{graphics}
\usepackage[latin1]{inputenc}
\usepackage{latexsym}
\usepackage{rotating}
\usepackage{hyperref}

\newcommand\be{\begin{equation}}
\newcommand\ba{\begin{eqnarray}}
\newcommand\ee{\end{equation}}
\newcommand\ea{\end{eqnarray}}

\begin{document}
%\title {Gauge Inflation, Baryogenesis and The Graceful Exit Problem}
\title {The Sterile Neutrino Field and Late Time Acceleration}
\author{Stephon H.S. Alexander}
%\email{sha3@psu.edu}
\affiliation{Department of Physics and Astronomy, Dartmouth College
Hanover, NH 03755} 

\date{\today}

%%%%%%%%%%%%%%%%%%%%%%%%%%%%%%%%%%%%%%%%%%%%%%%%%%%%%%%%%%%%%%%%%%%%%%%%%%%%%%%%%%%%%%%%%%%%%%
\begin{abstract}
\noindent 
We show that when the neutrino minimal standard model ($\nu MSM$) with a light sterile neutrino has a long range interaction between the neutrino vacuum charge density and a dark $U(1)$ gauge potential, late time acceleration is dynamically realized.  Moreover, the relation between the scale of dark energy, the weak and Planck masses derived by Arkani-Hamed et. al \cite{ArkaniHamed:2000tc} that addresses the coincidence problem arises naturally in this model.  Assuming that the cosmological constant is zero by some as yet known mechanism, the only fine tuning in our model is the mass of the sterile neutrino.  We show that perturbations of the vector fields are oscillatory and hence, there are no instabilities.  
\noindent 
\end{abstract}
%%%%%%%%%%%%%%%%%%%%%%%%%%%%%%%%%%%%%%%%%%%%%%%%%%%%%%%%%%%%%%%%%%%%%%%%%%%%%%%%%%%%%%%%%%%%%%

\pacs{11.25.Wx, 95.55.Ym, 04.60.-m, 04.80.Cc}
\maketitle

%%%%%%%%%%%%%%%%%%%%%%%%%%%%%%%%%%%%%%%%%%%%%%%%%%%%%%%%%%%%%%%%%%%%%%%%%%%%%%%%%%%%%%%%%%%%%%
\it{ Introduction} \rm While current cosmological data supports a $\Lambda CDM$ model, it plausible that the source of the current acceleration is due to an unknown fluid with a baryotropic index $w \sim -1$ or an IR modified version of general relativity.   Many models, mainly 
motivated by a fundamental scalar fields (such as quintessence) have been proposed to account for the late time acceleration.  On the other hand, most Dark energy models are challenged with the "why now" 
, or coincidence problem.   If the dark energy was due to a cosmological constant then we are faced with a fine tuning of the bare cosmological constant to one part in $10^{120}$.  If the dark energy is an unknown degree of freedom that is homogenous, we will have to understand the coincidence in the current epoch as to why $\rho_{DE} \sim 2\rho_{DM}$.  

It remains of interest to understand how dark energy may communicate or be related to the standard model of particle physics or its extension\cite{kam,Ando:2009ts}.  In this letter, we argue that a light sterile neutrino can also play an important role in the occurrence of late time acceleration.  For awhile, there has been an expectation that neutrinos and dark energy are connected, since the Atmospheric neutrino mass differences are proportional to the dark energy scale\cite{Fardon:2003eh,Alexander:2009yb,Bhatt:2009wb}.  This relationship was made concrete by relating the evolution of the dark energy through a Yukawa-like interaction between the quintessence field and a sterile neutrino; leading to a mass variation of neutrinos.  However, these models can exhibit instabilities on small scales \cite{Afshordi,Bean:2007ny}.    

The key to the onset of late time acceleration stems from a self consistent set of interactions between the sterile neutrino, a dark gauge field and the metric.  First, Sterile neutrinos, in an expanding space-time, experience a one-loop correction, which enhances the sterile neutrino charge density.  This charge in-turn sources a time-dependent amplification of a nearly constant gauge field mode.  As a result the gauge-fermion interaction which contributes to the energy-momentum tensor dominates and sources late time acceleration.  Furthermore, the perturbations of the neutrinos and gauge field exhibits oscillatory behavior and will be stable.

\noindent 
\it{The Model} \rm
\!\! In this model, we assume that there is a dark copy of abelian gauge field interacting with the sterile neutrino current whose symmetry group is $U(1)$ \footnote{Such dark sectors are quite common in String-Theory and Gravi-Weak Unification models \cite{AMS}}:  

\begin{eqnarray} \label{action1}
\!\!\!\!S= S_{\nu}\!+\!\!\int_{\mathcal{M}_4}\!\!\!\!\!\!\! d^4x\sqrt{-g} \Bigg[\frac{M_p^2\,R}{8\pi} - \frac{1}{4} F_{\alpha\beta}F^{\alpha\beta}\!  \Bigg]\,,
\end{eqnarray}
where $S_{\nu}$ is the covariant Dirac action
\be
S_{\nu}\!=\! \int_{\mathcal{M}_4} \!\!\!\!\!\! d^4x \sqrt{-g} \left(-i \overline{\nu}\,\slash\!\!\!\!\nabla\nu +c.c.+ M_{\nu}\bar{\nu}\nu + q\, \overline{\nu} \,\gamma^I e_I^{\nu} \nu \,A_{\nu}\right). \nonumber
\ee
 The tensor $F_{\mu \nu}\!=\!\partial_{[\mu}A_{\,\nu]}$ is the field strength tensor.  We denote as $q$ a dimensionless coupling constant. The neutrino current is $\mathcal{J}^\mu\!\equiv\! \bar{\nu} \,\gamma^I e_I^{\mu} \nu $, where $\nu$  is a sterile neutrino.
 
Using the relation $T^{\mu \nu} \!=\!  -\frac{2}{\sqrt{-g}} \frac{\delta \sqrt{-g} \tilde{\mathcal{L}}}{\delta g_{\mu \nu}}$, we find that the energy-momentum tensor is:
\begin{eqnarray} \label{tenso}
&T^{\mu}_{\nu}
&= {\rm Tr} \Big[ -q\,\delta^\mu_\nu A_\rho \mathcal{J}^\rho+F_\alpha^{\,\,\mu} F^{\alpha}_{\,\,\nu} \\
& &-\frac{1}{4} \delta^\mu_\nu g^{\alpha \rho} g^{\beta \sigma} F_{\alpha \beta} F_{\rho \sigma}- q\,A_{(\nu} \mathcal{J}^{\mu)} \Big].
\end{eqnarray}

We now wish to outline how cosmic acceleration can be driven by the purely time-like components of the gauge field $A_{0}$ and fermionic charge $\mathcal{J}_{0}$.  The condition for the gauge field is similar to scalar field driven quintessence, where one assumes that acceleration is driven by a spatially homogenous classical part plus quantum perturbations \footnote{From now on, we will be working in conformal coordinates $\{\eta,\vec{x}\}$.}, $\phi(x,\eta) = \phi_{0}(\eta) + \delta\phi(x,\eta)$. In our case
\be
 A_{\mu}(\eta,\vec{x}) = A^{(0)}_{\mu}(\eta) + \delta A_{\mu}(\eta,\vec{x}) \,.
\ee
Furthermore the spatial components of the fermionic current $J^i = 0$, vanishes due to the trace properties of the gamma matrices.  

The background solutions yields an energy-momentum tensor which sources a negative pressure equation of state, namely $A\!\cdot\! \mathcal{J}\!\equiv\! A_{\mu}\,\mathcal{J}^{\mu}$.  Perturbations are generated by the electromagnetic energy-density, $(  \vec{E}^2 +  \vec{B}^2 )/(2 a^4)$, which redshift as $a^{-4}$ since both electric and magnetic field propagate as radiation. The two sum up to the isotropic energy density 
\be
T_{00}=\!(  \vec{E}^2 +  \vec{B}^2 )/(2 a^4)+ q\, A\!\cdot\!\mathcal{J} = (\vec{\bar{E}}^2 +  \vec{\bar{B}}^2)/(2 a^4) + q\, \bar{A}_0 J^0\,,
\ee
 
 The accelerating phase is reached when the initial electric and magnetic fields (which depends on the perturbation $\delta A_{\mu}(\eta, \vec{x})$) are subdominant to the gauge-fermion energy $\vec{\bar{E_0}}^2 + \vec{\bar{B_0}}^2 <\!\!< q \bar{A}_0 J^0$.  This happens because the spatial gauge field is sourced by the spatial neutrino current which we will show to be vanishing in our scenario.  Moreover, during acceleration, the isotropic $A_{0}\!\cdot\!\mathcal{J_{0}}$ energy-density dominates over the anisotropic terms (off-diagonal terms) in the energy-momentum tensor, $A_{(\nu} \mathcal{J}^{\mu)}$.
 
We will find that the non-vanishing amplitude for the fluctuations of the background field $\delta A^{(0)}_{\mu}(\eta)$, are oscillatory which will not spoil isotropy provided that the initial fluctuations satisfy $|\delta A^{0}_\mu| <\!\!< A_\mu$.  Given these initial conditions at recombination, we shall now demonstrate that the coupled field equations indeed yield late time acceleration.

\it{The Gauge Fields and the Initial Conditions} \rm In what follows, we derive solutions to the field equations for the gauge field coupled to both the metric and the neutrino current.  Varying the action with respect to $A_{\mu}$
\ba \label{gy}
\eta^{\gamma \beta}\partial^{\alpha} \!F_{\alpha \beta}\! +\! a^4\,  \mathcal{J}^\gamma\!
\!=\!0\,,
\ea
where $\varepsilon^{\gamma \alpha \mu\nu}$ is the Levi-Civita symbol.  We seek to find self consistent solutions to all the above equations of motion, by working in conformal coordinates. 

As stated in the preceding section, the dark-energy domination begins with a time-dependent homogeneous background gauge field, $A^{(0)}_{\mu}(\eta) = (A^{(0)}_0(\eta),\,A^{(0)}_{i}(\eta))$. Moreover, the total gauge potential, including perturbations is
\begin{eqnarray}
& A_{\mu}(\eta,\vec{x}) = A^{(0)}_{\mu}(\eta) + \delta A_{\mu}(\eta,\vec{x}) \\
& =(A^{(0)}_0(\eta)+\delta A_{0}(\eta,\vec{x}),\,A^{(0)}_{i}(\eta)+\delta A_{i}(\eta,\vec{x}))\,.
\end{eqnarray}
Furthermore, we impose the Lorentz gauge, $\partial_\mu A^\mu=0$, then the $0$th component of (\ref{gy}) then yields the temporal component equation of motion 
\be  \label{A00}
\ddot{A}^{(0)}_0(\eta) = a^4 \mathcal{J}_0\,,
\ee

We can find an exact solution for the gauge field provided that the temporal fermion current is, $\mathcal{J}_0\sim J_0/a(\eta)$, which we will show 
later by computing the expectation value of the neutrino current.  We immediately find that the time-like gauge field grows proportional to the scale factor: $A^{(0)}_0(\eta)= \bar{A}_0\, a(\eta)$. For the spatial components of the background gauge field $\vec{A}^{(0)}$ the equation of motion is
\be \label{adri}
\vec{\ddot{A}}^{(0)}(\eta)=a^4 \vec{\mathcal{J}}\,.
\ee
We will find in the following section that $\vec{\mathcal{J}}=0$, resulting in $ \vec{\ddot{A}}^{(0)}=0$, which decays $\vec{A}^{(0)} \simeq \vec{c} + \vec{c}^{\,'}/a(\eta)$.  We assume that there is no background dark electric field filling the Universe, namely $\vec{c}=\vec{c}^{\,'}=0$.  

The equation of motion for the fluctuations around the background field components $A^{(0)}_\mu$, which we assume to be of infinitesimal order in some parameter $\lambda$, are now recovered to be from the variation of (\ref{gy})
\be \label{pert0}
\delta \ddot{A}_0(\eta, \vec{x})\, - \vec{\nabla}^2 \delta A_0(\eta,\vec{x})=0
\,,
\ee
for the $\delta A_0(\eta, \vec{x})$ component, whose solution is trivially found to be 
\be \label{solpert0}
\delta A_0(\eta, \vec{x})=\delta A_0 \, \exp (i k_0 \eta) \, \exp (-i \vec{k} \vec{x})\,.
\ee 
For $\delta \vec{A}(\eta, \vec{x})$ we find
\be \label{perti}
\delta\vec{\ddot{A}}-\nabla^2\delta\vec{A} =0 \,.
\ee
Without loss of generality, we can write the solution of (\ref{perti}) then cast the field equations in terms of Fourier modes 
\be
\delta \vec{A}(x,\eta) = \rm \int d^{3}k \,\sum_i \delta A(\eta,k)_{i}\, \epsilon_{i}(k) e^{ikx}\,. 
\ee
 The requirement that $\delta\vec{A}$ is traceless also ensures that $\vec{A}(\eta,k)$ is perpendicular to its direction of propagation. Our field equation for the gauge field then simplifies to
\be \label{fluxequation} 
\delta \ddot{A}(\eta,k)_i + k^2 \delta A(\eta,k)_i= 0 . 
\ee
The general solution for the spatial gauge field fluctuation%\footnote{We will be justified to not include the coupling of the superhorizon to the vector modes $V_{i}$, since spatial anisotropies on superhorizon scales are suppressed in this framework. We thank Justin Khoury for raising this point.} 
is found to be 
\be \label{evoluzione}
\delta A(\eta, \vec{x})_{i}=A_i^0\exp (i k_0 \eta) \, \exp (-i \vec{k} \vec{x}) \,.
\ee
So even the spatial perturbations are oscillatory for all wavelengths and will not experience any growth, and the electric and magnetic fields will not grow.  In the next section we recover the expression for the components of the fermionic current.

\it{Consistent Late Time Acceleration} \rm \noindent We now turn our attention to the Einstein Equations and seek an accelerating solution. 
In what follows we are going to employ a similar mechanism of space-time acceleration, used to obtain an epoch of cosmic inflation developed by \cite{Alexander:2011hz}.  In that model, the field content comprised of QED coupled to an axion in the visible sector of the Standard Model.  Our model differs in that it uses a dark $U(1)$ gauge-field coupled to sterile neutrinos and there are no axions.  
The $G_{00}$ component of the Einstein Equations gives the first Friedmann equation:
\ba \label{Fried}
3\frac{\dot{a}^2}{a^4}&=&\frac{8\pi G}{a^4} (\vec{E}^{2} + \vec{B}^2) + 8\pi G \, A_0\,\mathcal{J}^0\,.
\ea
The coupled system can be solved if the interaction term $A\!\cdot\!\mathcal{J}$ is nearly spatio-temporarily constant during inflation.  Since $A\cdot\mathcal{J}$ is constant we can solve the Friedmann equation (\ref{Fried}) subject to to our gauge field configuration $A_{\mu}(\eta,\vec{x}) = A^{(0)}_{\mu}(\eta) + \delta A_{\mu}(\eta,\vec{x})=(A^{(0)}_0(\eta)+\delta A_0(\eta, \vec{x}),\,A^{(0)}_{i}(\eta)+\delta A_{i}(\eta,\vec{x}))$, demanding that the initial conditions the electric and magnetic field to be $\vec{E}_{0}(\eta_{0})=\vec{B}_{0}(\eta_{0})=0$.  Therefore the Friedman equations reduces to
\ba
3\frac{\dot{a}^2}{a^4}&=  8\pi G \, A_0\,\mathcal{J}^0\,.
\ea
We then find an accelerating scale factor,
\be \label{dS}
a(\eta)=a_0\, [1- H (\eta-\eta_0) ]^{-1}\,,
\ee
where the Hubble parameter is $ M_p H\simeq \sqrt{(A \!\cdot\! \mathcal{J})_{\eta_0} }a_0 $. We can easily obtain the comoving scale factor $a(t)$ using the map $\partial\eta =\partial t/a$ (namely $a(t)=a_0\exp Ht$), which gives rise to an accelerating scale factor.  

\it{Fermionic Dynamics} \rm In the previous section, in order to get late time acceleration it is necessary that the neutrino current is not diluting as $\sim \frac{1}{a^{3}}$, but instead the VEV scales as $\langle J_{0} \rangle \sim \frac{1}{a}$.  Recently, Koskma and Prokopec evaluated the regularized fermion propagator self consistently in an FRLW background self-consistently\cite{Koksma:2009tc}.  
We can perform the proper contractions to obtain the expectation value fermion current and demonstrate that it indeed has the correct $\frac{1}{a}$ resdhifting. \cite{Alexander:2011hz}
%%
%%

%Using their results, we can perform the proper contractions to obtain the expectation value fermion current and demonstrate that it indeed has the correct $1\over a $ resdhifting. 

\noindent  Let us start by writing the free action for the fermionic field
\be
S=\int d^4 x \sqrt{-g} \left\{ \frac{i}{2} \left[ \,  \overline{\psi} \gamma^\mu \nabla_\mu \psi - \left( \nabla_\mu \overline{\psi} \right) \gamma^\mu \psi \right]  -m \overline{\psi} \psi   \right\}\,.
\ee

The equation of motion for $\psi$ are then
\be
i \gamma^\mu \nabla_\mu \psi(x) - m \, \psi(x) = 0\,.
\ee

The Feynman propagator $i G^{ab}_F (x, y)$ of the theory, satisfies the tree-level equation
\be
\sqrt{-g} \left( i \gamma^\mu \nabla_\mu -m \right)_x i G^{ab}_F (x, y)= \ i \delta^4 (x-y)\, 1\!\!1^{ab} \,.
\ee

The authors demonstrated that the fermion propagator is:\cite{Garbrecht:2006jm, Koksma:2009tc} 
\ba \label{fp}
&&i G^{ab} (x, y) = a(\eta_x) ( i \gamma^\mu \nabla_\mu + m) \frac{H^2}{\sqrt{a(\eta_x) a(\eta_y) }}\, \nonumber \\
&&\left[ i G_+(x,\,y) \frac{1+\gamma^0}{2} + i G_-(x,\,y) \frac{1-\gamma^0}{2}       \right]   \,, \nonumber\\
&& i G_\pm(x,\,y) = \frac{\Gamma(1\mp i \frac{m}{H} ) \, \Gamma(2 \pm i \frac{m}{H} )}{(4\pi)^2 \Gamma(2)}\!\! \nonumber \\
&&\phantom{a}_2 F_1\!\left( 1 \mp i \frac{m}{H}, \, 2\pm i  \frac{m}{H}, \, 2, 1-  \frac{\Delta(x,y)}{4} \right)\,, \nonumber\\
&& \Delta(x,y)= a(x) a(y) H^2 \Delta x^2\,, 
\ea
where $\varepsilon$ denotes an infinitesimal displacement on the imaginary time line.  The VEV of the sterile neutrino current components can be derived from the advanced or retarded propagator entering the definition of the Feynman propagator
\be
\langle   J^I  \rangle \simeq  \ {\rm lim}_{y \rightarrow x } \ G^{ab} (x, \, y)\gamma^I_{ba} 
\,.
\ee
Taking the limit in which the two space-time points coincide, $y\rightarrow x $, and the limit in which the infinitesimal displacement shrinks to zero, {\it i.e.} $\varepsilon \rightarrow 0$, we get
\ba \label{res}
G^{ab} (x, y)\!\simeq \!\frac{H^2}{16\pi^2}\!\left[ \! \frac{m^2}{H} \!\left( \frac{1+\gamma^0}{2}  \right)^{ab} \!\!\!\!\!-\! \frac{m^2}{H} \left(   \frac{1-\gamma^0}{2}  \right)^{ab} \! \right]\!,
\ea
in which we have used metric compatibility and expanded the hypergeometric function. 

We can now obtain the different components of the neutrino current by tracing (\ref{res}) with $\gamma^{I}$.   Because of the $\gamma^{0}$ factor in the propagator, the spatial current vanishes and the temporal current is non-vanishing.   We find that the temporal current is:
\be \label{J}
\langle   J^0  \rangle  \simeq  \frac{1}{4 \pi^2}\, m^2 H  \,,
\ee
and finally for the components of the fermion current $\langle \mathcal{J}^\mu \rangle $ 
\be
\langle \mathcal{J}^0 \rangle=\frac{\langle J^0\rangle}{a(\eta)}\simeq  \frac{1}{ 4 \pi^2} \, \frac{m^2 H}{a(\eta)}\,, \qquad \langle\mathcal{J}^i\rangle=0  \,.
\ee

We therefore conclude that the temporal current redshifts correctly so as to give late time acceleration.

\begin{figure}[tb] 
\begin{center}
 \vspace{-2mm}
\!\includegraphics[scale=0.4]{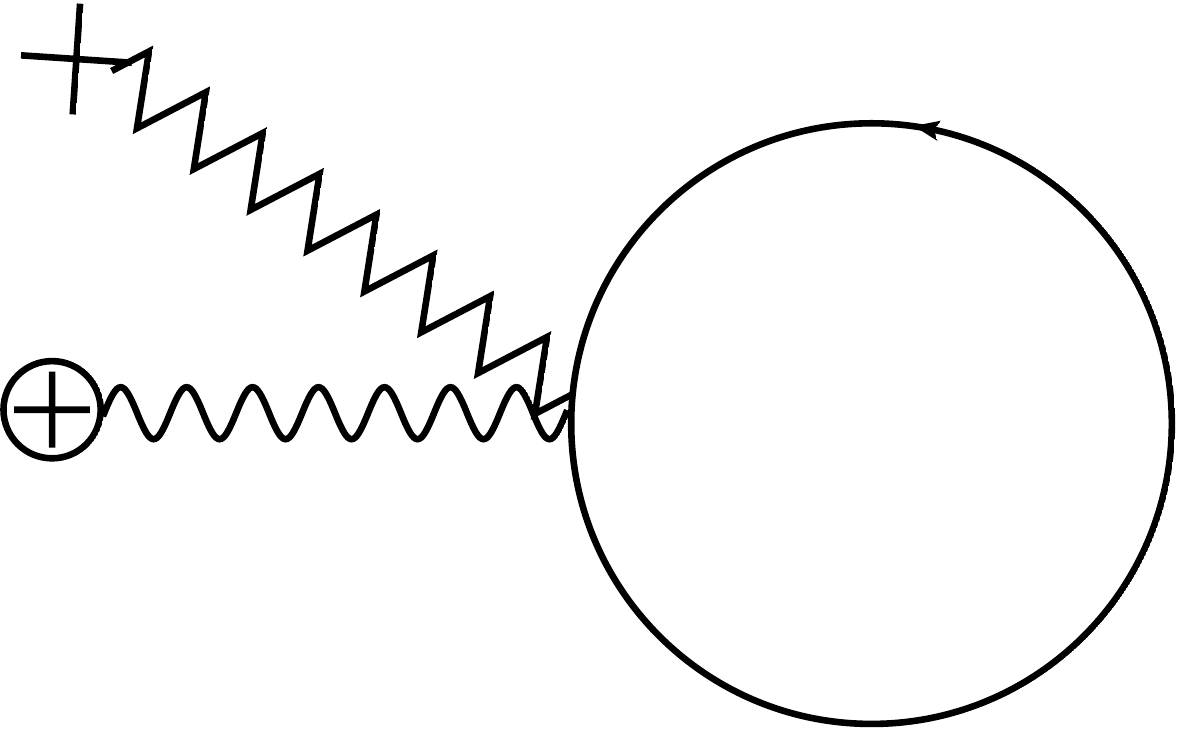}
\vspace{-2mm}
\caption{The Feynman diagram for the Sterile neutrino tadpole that is sourced by the background gravitational field.  The fermion loop exhibits a leading geometric contribution proportional to the Hubble parameter which sources the classical gauge field.  This is the dominant channel which contributes to the energy momentum tensor, leading to late time acceleration \label{f:cpratios}}.
\end{center}
\vspace{-2mm}
\end{figure}

\it{The Coincidence Problem} \rm We have shown that it is possible to generate late time acceleration from an interaction between a dark gauge potential with a sterile neutrino current in an expanding background.  We wish to show that the coincidence problem can be solved solely by knowing the mass of the Sterile neutrino.  In this sense our model has no-fine tunings provided we know the mechanism for generating and stabilizing the mass of the sterile neutrino.  Let us revisit the self consistent solutions of the coupled equations of motion; first the Friedmann equation:
\ba \label{Friedsec}
3H^{2}&=  8\pi G \, A_0\,\langle \mathcal{J}^0 \rangle \,.
\ea

The equation of motion (\ref{A00}) for $A(\eta)$ yields the solution $A(\eta)=\frac{a(\eta)J^{0}}{H^{2}}$ and for the VEV of the sterile neutrino charge, $\cal{J}\rm_{0} =\frac{m^{2}H}{a(\eta)}$.  Substituting the solutions of the gauge field and the fermion charge into the Friedmann equations gives the same relation derived by \cite{ArkaniHamed:2000tc} between the Hubble parameter and the mass of the sterile neutrino, 
\be 
H_{0} \simeq \frac{M_{\nu_{s}}^{4}}{M_{pl}^{2}}\,. 
\ee   
If we take the Hubble parameter to be the value of the dark energy today, $H_{0}= \frac{M_{DE}^{2}}{M_{pl}} \simeq 10^{-42} GeV \, $
we determine the mass of the Sterile neutrino, which is found to be millielectron-volts, $M_{\nu_{s}} \sim 10^{-3} eV$.  In otherwords, if the mass of the sterile neutrino is miliectron-volts then then we dynamically get the observed late time acceleration.  

This model can be accommodated by the minimal extension of the standard model, $\nu MSM$, with three right-handed singlet sterile neutrinos \cite{AS,DP} :
\be \cal{L}\rm = \cal{L}\rm_{SM} + i\bar{N}_{I}D_{\nu}\gamma^{\mu}N_{I} - \left( F_{\alpha I}\bar{L}_{\alpha}N_{I}\phi -\frac{M_{I}}{2}\bar{N}^{c}N_{I} + h.c \right) \,,
\ee

where $F_{\alpha}$ are new Yukawa couplings, $L_{\alpha}$ are left-handed leptons and $\alpha$ labels the generations.  Our model can be accommodated by the $\nu MSM$ if one of the right handed neutrino is gauged under a dark $U(1)$ local phase transformation.  Since the sterile neutrinos couple to active neutrinos via mixing, they must live longer than the lifetime of the universe if they are to source late time acceleration.  The main decay channel is $N \rightarrow \bar{\nu}\nu\nu$ or $N \rightarrow \bar{\nu}\bar{\nu}\nu$, for wich the decay time is:
\be 
\tau = 5\times10^{23}s \, \left(\frac{M}{1keV}\right)^{-5}\left(\frac{\theta^{2}}{10^{-5}}\right)^{-1}\,.
\ee
Given that the age of the universe is $10^{17}\,s$, we find that for a mass of $M \sim 10^{-3} eV$ the neutrinos are stable against decay.

\it{Conclusion} \rm In this letter we connect the interaction between a light sterile neutrino and a dark gauge potential as the cause for the onset of late-time acceleration.  We demonstrate that a long range time-like vector potential sourced by the sterile neutrino vacuum charge density can generate the correct magnitude in energy density to drive late time acceleration.  The coincidence problem has only one fine-tuning, namely the neutrino mass.  In past works, various authors have considered the astrophysical consequences of dark radiation arising from the decay of a dark fermion into a dark gauge field\cite{kam2,Fan:2013tia}.  While our model posits a dark $U(1)$ sector, our gauge field is not radiative, since the only component of the gauge field $A_\mu =(A_{0}(t),0,0,0)$ is time-like and all spatial components are vanishing.   We have demonstrated that all spatial fluctuations of the gauge fields are oscillatory and will redshift away.

It is also interesting that the $\nu MSM$ can accommodate three sterile neutrinos.  The heaviest sterile state on the scale of $M \sim 10^{12} GeV$ is responsible for the smallness of the active neutrino masses and potentially leptogenesis due to rapid decay of the heavy sterile into leptons.  Furthermore, a $KeV$ scale sterile can be a dark matter candidate since it is stable against decay over the lifetime of the universe.  One may wonder if these other neutrinos could contribute to acceleration of the universe, and this would cause a problem for this mechanism since the Hubble rate has a quartic dependence on the mass of the neutrino.  We may be able to circumvent this problem if the heavier neutrinos couple to massive gauge fields whose interactions will be screened at large scales.  We therefore must assume that either the other heavier sterile states do not couple to a dark $U(1)$ gauge field or if they do, the gauge field must only have a short range interaction. We will pursue this issue in a forthcoming paper \cite{stephon} in the context of the Pati-Salam model.

From an observational point of view ultra light sterile neutrino may mix weakly with both atmospheric and solar neutrinos and might give rise to observable eects in future neutrino data \cite{DP}.  Finally, ultra-light sterile neutrinos have been invoked as an explanation for the extra radiation observed in the Universe \cite{S}.  It will be interesting to study the possibility of mixing between the dark energy and radiative states to make observational contact with CMB experiments.%\cite{stephon} 
In a companion paper \cite{AMS2}, we will address late time acceleration using a non-abelian dark gauge sector. %%

\section*{Acknowledgement}
\noindent
I especially thank Antonino Marciano for his useful feedback during the course of this project and for reading a draft of this paper.  I also thank Robert Caldwell for his critical feedback.

\end{document}